\documentclass{article}
\usepackage[T1]{fontenc} 
\usepackage[utf8]{inputenc} 
\usepackage{ismir,amsmath,cite,url}
\usepackage{graphicx}
\usepackage{color}
\usepackage[bookmarks=false,hidelinks]{hyperref}
\usepackage{lineno}
\usepackage{subcaption}
\usepackage{enumitem}
\usepackage{tabularx}
\usepackage{arydshln}
\usepackage{multirow}

\usepackage[nolist]{acronym}

\acrodef{omv}[OMV]{Official Music Video}
\acrodefplural{omv}[OMV]{Official Music Videos}
\acrodef{se}[SE]{Structural Events}
\acrodef{mir}[MIR]{Music Information Retrieval}

\title{Is there a "language of music-video clips" ? \\ A qualitative and quantitative study}

\threeauthors
  {Laure Prétet} {LTCI, Télécom Paris \\ Bridge.audio, Paris, France}
  {Gaël Richard} {LTCI, Télécom Paris \\ IP Paris, France}
  {Geoffroy Peeters} {LTCI, Télécom Paris \\ IP Paris, France}

\begin{document}

\maketitle


\begin{abstract}
Recommending automatically a video given a music or a music given a video has become an important asset for the audiovisual industry - with user-generated or professional content.
While both music and video have specific temporal organizations, most current works do not consider those and only focus on globally recommending a media.
As a first step toward the improvement of these recommendation systems, we study in this paper the relationship between music and video temporal organization.
We do this for the case of official music videos, with a quantitative and a qualitative approach.
Our assumption is that the movement in the music are correlated to the ones in the video.
To validate this, we first interview a set of internationally recognized music video experts.
We then perform a large-scale analysis of official music-video clips (which we manually annotated into video genres) using MIR description tools (downbeats and functional segments estimation) and Computer Vision tools (shot detection).
Our study confirms that a "language of music-video clips" exists; i.e. editors favor the co-occurrence of music and video events using strategies such as anticipation. 
It also highlights that the amount of co-occurrence depends on the music and video genres.
\end{abstract}


\section{Introduction}
\label{sec:introduction}


Each day, an ever-growing quantity of videos is created  by professionals (for advertisement, movies, series, etc) and individuals (for Instagram, TikTok, YouTube, etc).
Finding an appropriate soundtrack to emphasize the video content is therefore a common exercise, which can be time-consuming if done manually.
This explains the success of commercial systems such as \href{https://www.matchtune.com/}{MatchTune} or of research papers such as ``Look, Listen and Learn'' \cite{Arandjelovic2017LookLearn}.
While such systems are very good at recommending music based on the video content, the temporal synchronization between both modalities is rarely taken into account.
In order to develop synchronization-aware recommendation systems, some domain knowledge is required on how the synchronization is performed in real videos that feature music.
In this work, we attempt at bridging this knowledge gap by performing a fine-grained cross-modal analysis of the synchronization between audio and video content.
We hypothesize that better understanding professionally produced music videos helps designing better models for music-video synchronization.
This has applications in automatic music-video recommendation \cite{Gillet2007OnVideos, Wang2012TheVideo, Kuo2013BackgroundAnalysis, Lin2015EMV-matchmaker:Generation, Lin2017AutomaticEditing} and generation \cite{HUA2004AutomaticAnalysis, Wang2006FullyComposition, Wang2007GenerationCues}.

Temporal structure (at the beat, bar or functional segment level) is one of the dominant characteristics of music. 
For this reason, its automatic estimation has received a lot of attention in the \ac{mir} community \cite{DownieMusicEXchange.}.
Temporal structure in video (cuts, scenes, chapters) has similarly received a lot of attention in the Computer Vision community (for example with the goal of creating video summary \cite{Aner2002VideoClustering}).
Our fine-grained analysis will be using these structural elements.

Our cross-modal analysis could be performed on any type of video that features a musical soundtrack (eg commercials, movies).
We focus here on the special case of of \acp{omv}.
We call \ac{omv} an audiovisual document where the audio part consists in a music track, and which aims at promoting said track and its performing artists.
As a result, the music track is generally the only source of audio in \acp{omv}s.
This makes \acp{omv}s good prototypes for a study on music-video synchronisation.
We do not consider user-generated videos, because we assume that analyzing professionally produced \acp{omv}s is more likely to provide reusable insights.

In the specific case of \acp{omv}s, the editing team will often arrange the video rushes based on the structure of the music track \cite{Schindler2016HarnessingRetrieval}.
In some cases, the music track can also be adapted from the studio version for narrative purposes.
Therefore, music and video structure are de facto associated.
However, the level of synchronicity is not always the same, depending on the considered \ac{omv}.
This is not only due to artistic choices but also depends on the music genre and video genre, as we will see in our study.

\textbf{Proposal and paper organization.}
%
In this paper, we study the relationship between music and video temporal organization using a qualitative and a quantitative  approach.
The \textit{qualitative} study is based on a set of interviews with three renowned specialists of official music videos.
We interview them in order to find out if and how they consider the relationship between music and video structure in their work.
The \textit{quantitative} analysis is based on a detailed analysis of music and video structural events in \ac{omv} using \ac{mir} and Computer Vision tools.
The \ac{omv}s correspond to a subset of the Harmonix dataset~\cite{Nieto2019TheMusic}.
We study specifically the relationship between the duration of music and video segments and between the positions of their respective boundaries.
We highlight the dependency of those according to the \ac{omv} music and video genre (for which we annotated the data).

The paper is organized as follows.
Section \ref{sec:soa} discusses the related literature.
Section \ref{sec:itw} describes the qualitative study and summarizes the interviews of three music video experts: Jack Bartman (composer), Alexandre Courtès (director) and Maxime Pozzi (editor).
Section \ref{sec:analysis} describes the quantitative study: the dataset creation (\ref{ssec:dataset}), the analysis of the music and video segment duration (\ref{duration}) and of the music and video segment position (\ref{position}).
Section \ref{sec:conclusion} concludes and discusses the perspectives of this work.


\section{Related work}
\label{sec:soa}

\subsection{Music-Video Synchronization: A Short Review}

Music supervision is an industry that aims specifically at synchronizing music to video.
Music supervision experts are dedicated to proposing the best soundtrack to all types of videos, ranging from commercials to movies and series.
As of today, this recommendation and synchronization work still features a large amount of manual work.
Inskip et. al. \cite{Inskip2008MusicRetrieval} interviewed music supervision experts and described their workflow.
The authors mention that "the clearest briefs appear to be moving images", suggesting that other types of data (emotion, textual description, reference soundtracks) are not necessary to perform the task.

At the same period, Gillet et al. \cite{Gillet2007TranscriptionAudiovisuelles} proposed a system that can automate part of the music supervision task.
Their system relies on the synchronization of music structure (onsets and functional segments) and video structure (motion intensity and cuts) to perform music-video recommendation, without external data. Yang \cite{RuiduoYang2004MusicObservation} and Mulhem \cite{Mulhem2003PivotMixing} proposed similar approaches.

More recently, Alexander Schindler gathered a dataset of \acp{omv}s (the Music Video Dataset) and performed an in-depth analysis of this specific media \cite{Schindler2019Multi-ModalAnalysis}.
In \cite{Schindler2019OnVideos}, Schindler and Rauber explain how shot boundary detection is an ill-defined task in music videos, as shot transitions are used in a complex and artistic way.
By analyzing the clips, they observe that the music videos present characteristic editing styles (number of shots per second, types of transition) for certain music genres or moods. 
But they do not quantify this correlation.
In \cite{Schindler2016HarnessingRetrieval}, the same authors analyze the correlation between visual contents (objects present in the scene) and music genre.
For example, cowboy hats are almost systematic in country music videos. 

In our study, we propose a joint approach.
We analyze the correlation between the music/video \textit{structure} and music/video \textit{genres}.

\subsection{Audiovisual Structure Estimation Tools}

Our quantitative study (Section \ref{sec:analysis}) relies both on MIR to analyze the music structure and on Computer Vision to analyze the video structure.
More specifically, we estimate the downbeat positions, functional segments and shot boundaries from the \ac{omv} of our dataset.
In the following, we describe the tools we have used for our analysis.

Downbeat tracking is a popular \ac{mir} task \cite{Jia2019DeepReview}.
As a result, several ready-to-use libraries are available to estimate downbeat positions from audio files \cite{Bock2016Madmom:Library, Bogdanov2013ESSENTIA:Analysis}.
The state-of-the-art algorithm of Böck et al. \cite{Bock2016JointNetworks.} consists in two steps.
First, a \texttt{RNNDownBeatProcessor}, which relies on multiple Recurrent Neural Networks (LSTMs), estimates jointly beat and downbeat activation functions.
The output of the neural networks represents the probability of each frame of being a beat or downbeat position.
These activation functions are then fed as observations to a \texttt{DBNDownBeatTrackingProcessor}, which relies on a Dynamic Bayesian network (DBN).
The DBN outputs the beat positions of highest likelihood, along with their position inside the bar.

At a larger timescale, the automatic detection of boundaries between functional segments (choruses, verses and bridges) has also received a lot of attention from the MIR community.
The Ordinal Linear Discriminant Analysis (OLDA) algorithm by McFee et al. \cite{McFee2014LearningAnalysis} relies on supervised learning to perform this task.
This method adapts the linear discriminant analysis projection by only attempting to separate adjacent segments.
Then, the obtained features are clustered with a temporal constraint: only similar successive segments are merged together.

Similar to music, videos can be divided into segments of various duration, from shots to scenes to chapters and longer sequences.
In this study, we focus on a segmentation into shots.
The TransNet system \cite{Soucek2019TransNet:Transitions}, by Souček et al., is a Convolutional Neural Network which employs dilated 3D convolutions and which is trained in a supervised way on a shot boundary detection task.

\vspace{-0.3cm}
\section{Qualitative analysis: interviews}
\label{sec:itw}

\subsection{Methodology}

In order to gather intuition on the synchronization of music and video, we conducted a series of semi-structured face-to-face interviews.
We selected three music video experts from different professions: composition, direction and editing.
Following Inskip et. al. \cite{Inskip2008MusicRetrieval}, we selected the respondents using a snowball sampling technique.

Interviews were performed using the Zoom video conferencing software, lasting up to one and a half hours. The interviews were transcribed manually by the researcher, and transcripts were sent back to the respondents for validation.
Areas of discussion included the participant's day-to-day workflow and technical tools, their interactions with the other professions of the industry, and their opinion on example music videos prepared by the researcher.


\subsection{Interviews Summary}

\subsubsection{Jack Bartman, Composer}
\label{ssec:itw_jack}

As a composer (for commercials such as Nike, Apple or UbiSoft), Bartman has to adopt both a global and a precise local approach: the content of the music has to match the visual atmosphere, and its temporal structure must be aligned both globally at clip level and locally at frame level.
In some cases, the editing follows the structure of the music.
But in other cases, typically for advertisement, it is the opposite, and the composer has to adapt the music to an existing movie.
Most of the time, when music has to be edited on an existing movie, the slicing operation is privileged.

\textit{"Slicing can happen at unconventional moments, like the first or last beat of a bar! I simply add a sound effect to make it work."}

Time stretching and accelerations can be employed too, but are far less usual.
Bartman stresses that synchronizing cuts to audio events is especially important around emotional climaxes of the video.
Finally, for some projects, an exact synchronization is not the golden rule:

\textit{"This year, I worked on a short movie about psychological aspects of the Covid-19 lockdown.
After getting used to an imperfectly synchronized mockup soundtrack, the director did not want to use the final version, as the mockup would better suit the intended "madness" atmosphere".}


\subsubsection{Alexandre Courtès, Director}
\label{ssec:itw_alex}

As a director (such as for U2, Phoenix, Cassius, Franz Ferdinand or Jamiroquai), Courtès generally has a lot of freedom when it comes to the temporal organization of a music video.
Directors often come up with their own concept and they have little constraint about the content of the video.
At large temporal scale, their mission is to emphasize the music climaxes by the appropriate video content.

\textit{"The music video will often show a performance, so it is similar to a musical comedy: it has to feature costumes, chapters, sets, acts."}

Directors are not responsible for placing the cuts, but they can introduce diversity in the video transitions (explosions, large objects passing in front of the camera; see \cite{Schindler2019OnVideos} for a more exhaustive list).

\textit{"Cuts have to follow the music's rhythm, even though they might not always co-occur with beats."}


\subsubsection{Maxime Pozzi, Editor}
\label{ssec:itw_maxime}

As an editor (such as for Rihanna, Taylor Swift, Foals or Woodkid), Pozzi has to combine both a local, frame-level approach to the design of a global emotional trajectory.
    
\textit{"Editors and musicians have a similar job, we all want the same thing: rhythm, narration, climaxes."}

\begin{figure}
    \centering
    \includegraphics[width=\columnwidth]{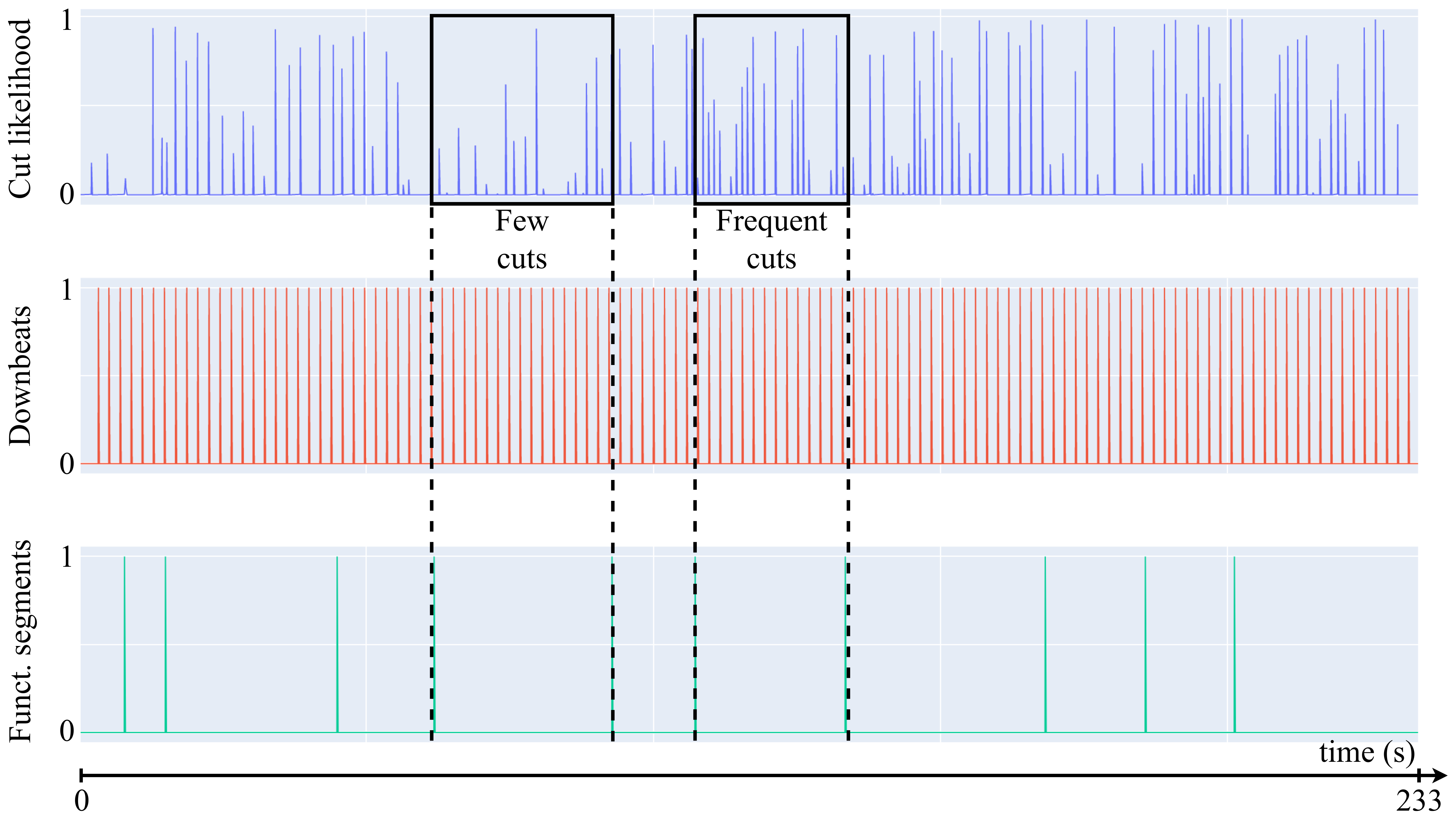}
    \caption{Audiovisual structure of Katy Perry, \textit{Firework}, full clip. Horizontal axis: time. Cuts: TransNet estimates. Downbeats: Madmom estimates. Music functional segments: OLDA estimates.}
    \label{fig:firework}
\end{figure}

For chorus and verses, the editing will follow the rhythm and typically accelerate near climaxes. During bridges, it will often be slower and poetic. 
This can be illustrated for example by Katy Perry's \href{https://www.youtube.com/watch?v=QGJuMBdaqIw}{\textit{Firework}} music video (Figure \ref{fig:firework}).
In this clip, we can see some functional segments where cuts happen very frequently (several times in each bar) and segments where they happen less frequently, for example on the downbeats only.

\begin{figure}
    \centering
    \includegraphics[width=\columnwidth]{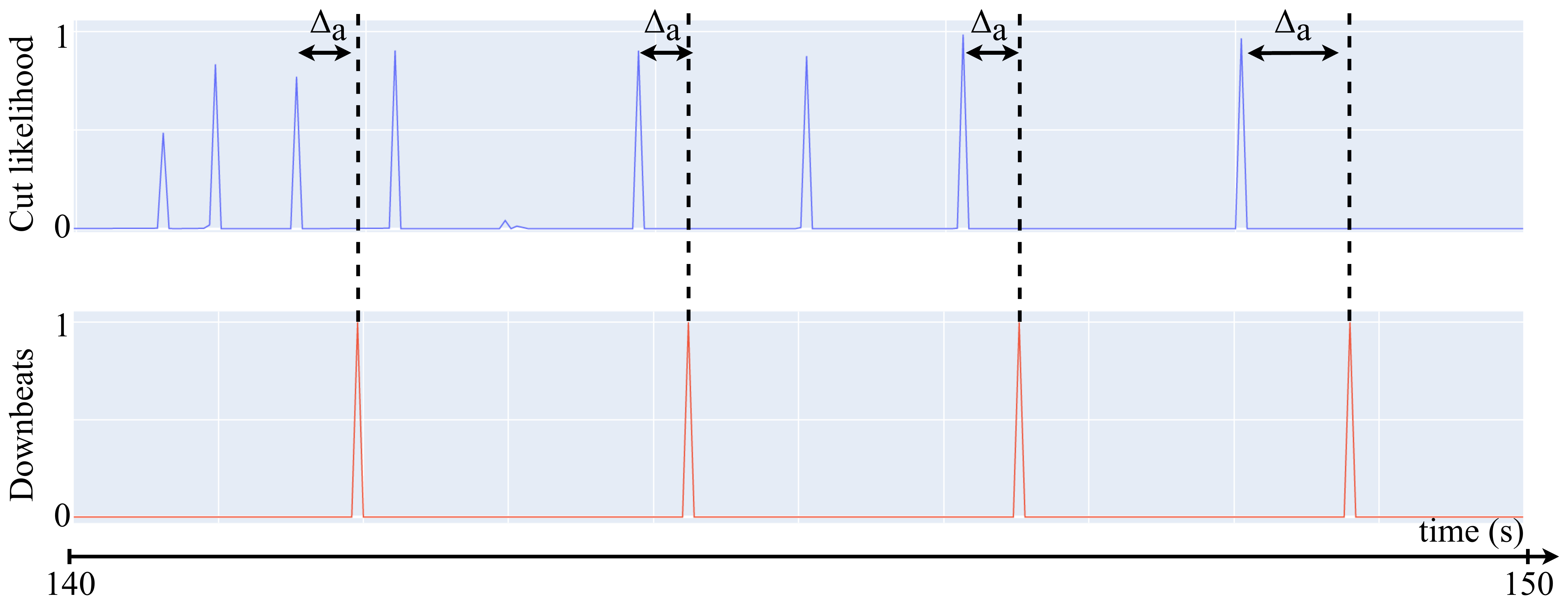}
    \caption{Audiovisual structure of Adele, \textit{Rolling in the deep}, timestamps 02:20 to 02:30. Horizontal axis: time. Cuts: TransNet estimates. Downbeats: Madmom estimates. $\Delta_a$: anticipation of cuts with respect to downbeat.}
    \label{fig:rollinginthedeep}
\end{figure}

Editing can be used as an element of narration.
For example, in Adele's \href{https://www.youtube.com/watch?v=rYEDA3JcQqw}{\textit{Rolling in the deep}} music video, starting at timestamp 02:20, the cuts are systematically placed just before the downbeat (see Figure \ref{fig:rollinginthedeep}).

\textit{"Off-beat cuts are used to create dynamics: to surprise the viewer, and illustrate the music's emotional climax. It makes the video direction appear more "indie" as well, this can be required by the performing artists."}

\subsection{Summary}

These three interviews provide us with a series of intuitions and hypotheses about the way audio and video are synchronized in music videos.
First, musical structure such as  chorus and verses are taken into account when directing a music video. 
Second, audio events such as rhythm, beat and downbeat are taken into account when editing a music video.
Finally, according to the desired atmosphere, the audio and video structural events can be more or less perfectly synchronized.


\section{Quantitative Analysis}
\label{sec:analysis}

\subsection{Methodology}
\label{ssec:method}

In the following, we conduct a set of quantitative experiments on how the \ac{se} of the music and of the video are synchronized in time.
We do so using Official Music Videos (OMVs).
We therefore first collect a dataset of \acp{omv}s, along with music and video genre annotations (Section \ref{ssec:dataset}).
For each of them we use \ac{mir} tools to estimate music \ac{se} (downbeats and functional segments) and Computer Vision tools to estimate video \ac{se} (shot boundaries).
In our first experiment, we study the correlation between the duration of the shots and the various musical \ac{se}s  (beat and bar duration).
In our second experiment, we study the temporal co-occurrence of the shot boundaries and the various musical \ac{se}s (bar and functional segment boundaries).
We analyze the results of those for each music genre and each video genre.

\subsection{Dataset}
\label{ssec:dataset}

For our quantitative study, we consider a subset of the Harmonix dataset \cite{Nieto2019TheMusic}.
Harmonix was initially released for automatic estimation of beat, downbeat and functional music segments.
It features popular (mostly hits) Western music tracks for which there is a high probability of having an associated music video.
%
%
From the list of 1,000 YouTube video links provided, 899 were successfully retrieved, of which 40\% contained only still images and 2.4\% were duplicates. 
As a contribution of this work we provide the list and URLs of the remaining 548 \ac{omv}s as well as the genre annotations described below\footnote{Our list is accessible at: \href{https://gitlab.com/creaminal/publications/ismir-2021-language-of-clips/-/blob/master/video_genres.csv}{https://gitlab.com/creaminal/publications/ismir-2021-language-of-clips/-/blob/master/video\textunderscore genres.csv.}}.

\begin{figure}
    \centering
    \includegraphics[width=1\columnwidth]{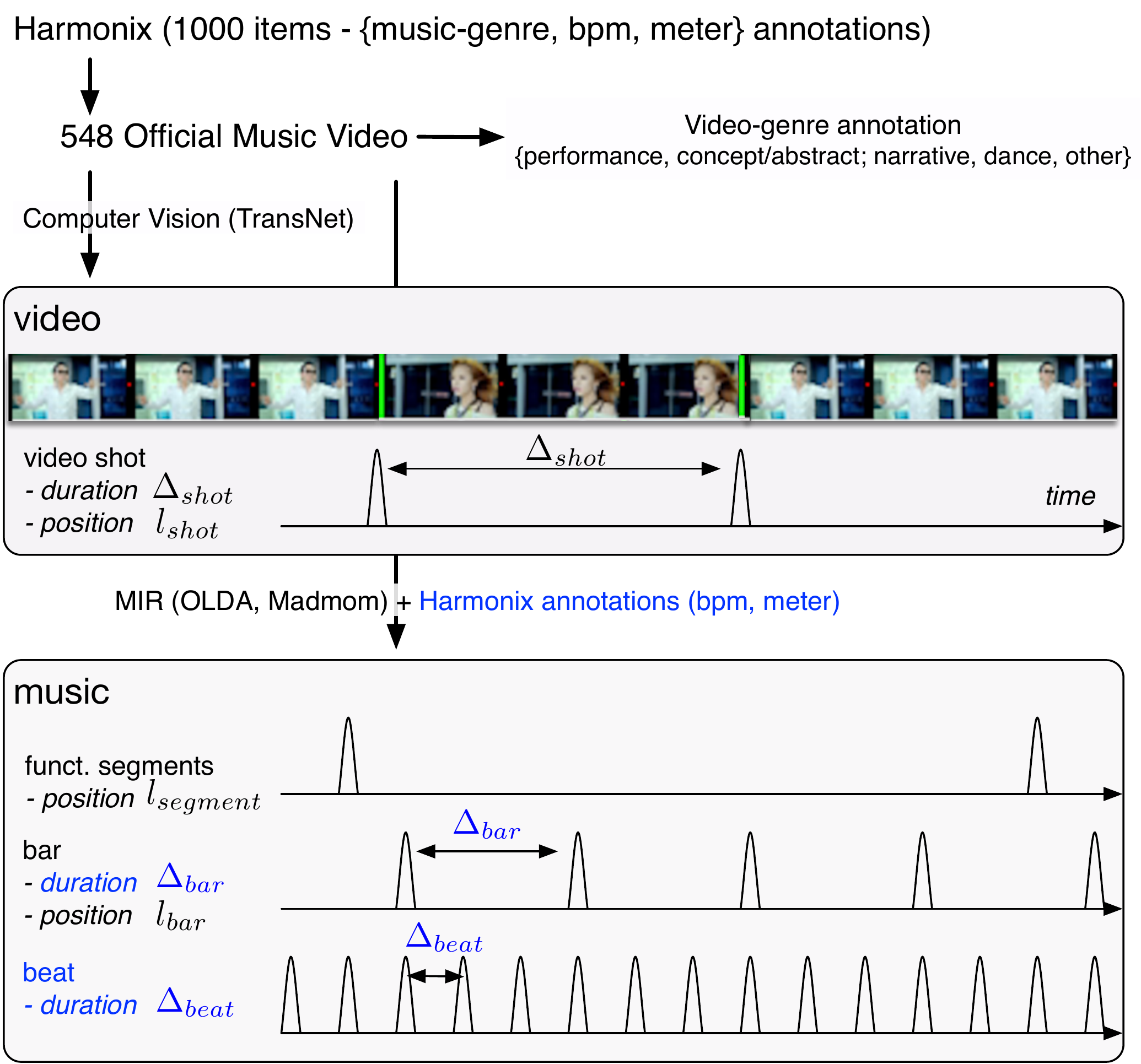}
    \caption{Schematic view of the different audiovisual structural events considered: shots ($\Delta_{shot}, l_{shot}$), functional music segments ($l_{segment}$), bars/downbeats ($\Delta_{bar}, l_{bar}$) and beats ($\Delta_{beat}, l_{beat}$). 
    Illustration music video: Psy, \textit{Gangnam Style}.}
    \label{fig:av_structure}
\end{figure}

\subsubsection{Annotations into \acl{se}}

We consider here two types of \acf{se}: those based on the music content -audio-, and those based on video content -image frames over time (see Figure \ref{fig:av_structure}).

\textbf{Music \ac{se}.}
We consider three types of music \ac{se}s.
At the smallest temporal scale we consider the beats and downbeats; at the largest temporal scale we consider the functional music segment boundaries (between the verses, bridges, choruses).
Harmonix features a set of manual annotations\footnote{into functional segments, downbeat and beat.}.
However, these annotations correspond to studio versions of the tracks which can, in some cases, be largely different from the version used in the \ac{omv}.
For this reason, we only used the annotations into bpm and meter of the Harmonix dataset to get the beat duration $\Delta_{beat}=\frac{60}{bpm}$ and bar duration $\Delta_{bar}=4 \mbox{ or }3 \Delta_{beat}$ (which is computed as a multiple of the bar duration using the time signature).
For the downbeat positions, we used the algorithm of Böck et al. \cite{Bock2016JointNetworks.}, implemented in the Madmom library \cite{Bock2016Madmom:Library}.
In the following, we denote by $l_{bar}$ the list of downbeat positions for a given track.
For the functional music segments, we used the implementation of OLDA from the MSAF library~\cite{Nieto2016SystematicResearch.}. 
In the following, we denote by $l_{segment}$ the list of boundary positions between the segments for a given track.
For our dataset, the average duration of functional music segments is 19.73~s. and the average bar duration is 2.30~s.

\textbf{Video \ac{se}.} 
We consider only the least ambiguous video \ac{se}, the shot boundaries (or cuts).
To detect boundaries between shots, we use the TransNet system \cite{Soucek2019TransNet:Transitions} and the associated library, available on GitHub\footnote{\href{https://github.com/soCzech/TransNet}{https://github.com/soCzech/TransNet}}.
The TransNet output is a continuous function of time $f_{shot}(t) \in [0,1]$ representing the likelihood of a boundary at time $t$.
$f_{shot}$ has a sampling rate of 25 Hz.

Also, for each \ac{omv}, we compute the histogram of its shot duration.
We do so by first estimating the list of shot boundary positions $l_{shot}$ by thresholding $f_{shot}(t)$ with $\tau=0.5$.
The resulting segments have an average duration $\Delta_{shot}$ of 4.76s.
We then compute the histogram of these durations.
We denote by $\Delta_{shot}^{\max}$ the position of the maximum of this histogram (in seconds).

We sum up the various \ac{se} in Table \ref{tab:notations}.

\begin{table}[h!]
    \centering
    \footnotesize
    \caption{Notation associated to each SE considered.}
    \label{tab:notations}
    \begin{tabular}{|l|l|l|}
        \hline
        \textbf{Music} & \\
        \hline
        genre & & Harmonix annotations \\
        \hdashline
        funct. segments positions & $l_{segment}$ & OLDA/MSAF \\
        bar duration & $\Delta_{bar}$ & Harmonix annotations \\
        bar/downbeat positions & $l_{bar}$ & Madmom \\
        beat duration & $\Delta_{beat}$ & Harmonix annotations \\
        \hline
        \textbf{Video} & \\
        \hline
        genre & & Manual annotations \\
        \hdashline
        shot boundary probability & $f_{shot}(t)$ & TransNet \\
        shot boundary positions & $l_{shot}$ &  \\
        most common shot duration & $\Delta^{\max}_{shot}$ &  \\
        \hline
    \end{tabular}
\end{table}

\subsubsection{Annotations into genre}

We consider both the genre associated to the music and the one associated to the video.

\textbf{Music genre.}
While still controversial in its exact definition~\cite{Hennequin2018AudioTags}, music genre is a convenient way to describe musical content. 
For this reason, it has been and it is still a widely studied topic\footnote{It has \href{https://www.music-ir.org/mirex/wiki/2020:Audio_Classification_(Train/Test)_Tasks}{dedicated challenges} \cite{Defferrard2018LearningAudio}, and large datasets featuring hundreds of categories \cite{Bertin-Mahieux2011TheDataset, Gemmeke2017AudioEvents, Defferrard2017FMA:Analysis}.}.
For our experiments, we use the music genre annotations provided by the Harmonix dataset metadata.

\textbf{Video genre.}
Video genre classification is a much less studied topic.
Existing studies focus on a much smaller sets of video genres \cite{You2010AAnalysis, Varghese2019ARelevance, Choros2018VideoShots}.
Only Gillet et al.~\cite{Gillet2007OnVideos} and Schindler~\cite{Schindler2019Multi-ModalAnalysis} studied the case of \ac{omv}s and there is no consensus on their taxonomy of video genres. 
There is also no annotated dataset for this task.

We merge \cite{Gillet2007OnVideos} and \cite{Schindler2019Multi-ModalAnalysis} to obtain a set of 5 video categories and and a corresponding single-label dataset.
Maxime Pozzi, a professional music video editor, validated our taxonomy during our preliminary interview (see part \ref{ssec:itw_maxime}).
One author then manually annotated all 548 video clips of Harmonix into the five following video genres:

\begin{itemize}[leftmargin=4mm, parsep=0cm, itemsep=0cm, topsep=0cm]
    \item \texttt{Performance videos (P):}  The artist or band are presented performing the song. 74 videos; example: Iron Maiden, \textit{Powerslave}.
    \item \texttt{Concept/Abstract videos (C):} The video illustrates the music metaphorically via a series of abstract shots related to semantics or atmosphere of the song. 227 videos; example: Lady Gaga, \textit{Poker Face}.
    \item \texttt{Narrative videos (N):} The music video has a strong narrative content, with identifiable characters and an explicit chronology. 160 videos; example: Taylor Swift, \textit{Teardrops on My Guitar}.
    \item \texttt{Dance videos (D):} Artists present a rehearsed dance choreography in sync with the music. 62 videos; examples: Sean Paul, \textit{Get Busy}.
    \item \texttt{Other (O):} Other types of music videos, including lyrics videos, animated music videos, etc. 25 videos; example: Train, \textit{Hey, Soul Sister}.
\end{itemize}

\subsection{Experiments}
\label{ssec:experiments}

We hypothesize that the music structural events play an important role for the placement of cuts during the video editing.
We check this assumption by measuring:
\begin{itemize}[leftmargin=4mm, parsep=0cm, itemsep=0cm, topsep=0cm]
    \item if their segment duration are correlated in Section \ref{duration};
    \item if their position co-occur in Section \ref{position}.
\end{itemize}
According to Gillet \cite{Gillet2007OnVideos}, the performance of alignment-based music-video recommendation systems are strongly correlated to the video genre. 
We therefore differentiate our results by music and video genre.

\subsubsection{Comparison between events duration}
\label{duration}

Our first experiment aims at evaluating to which extent the musical and video events have similar durations.

To measure this, we compare $\Delta_{shot}^{\max}$ (the most common shot duration) with the beat duration $\Delta_{beat}$ and bar duration $\Delta_{bar}$ obtained from the Harmonix annotations.
When $\Delta_{shot}^{\max}$ is close to $\Delta_{bar}$, this indicates that a systematic change of shots occurs with the same speed as the bar changes. 
This however does not mean that the changes occur simultaneously (we study this in Section \ref{position}).

\begin{figure}[t]
    \begin{subfigure}{\columnwidth}
      \centering
        \includegraphics[width=0.9\columnwidth]{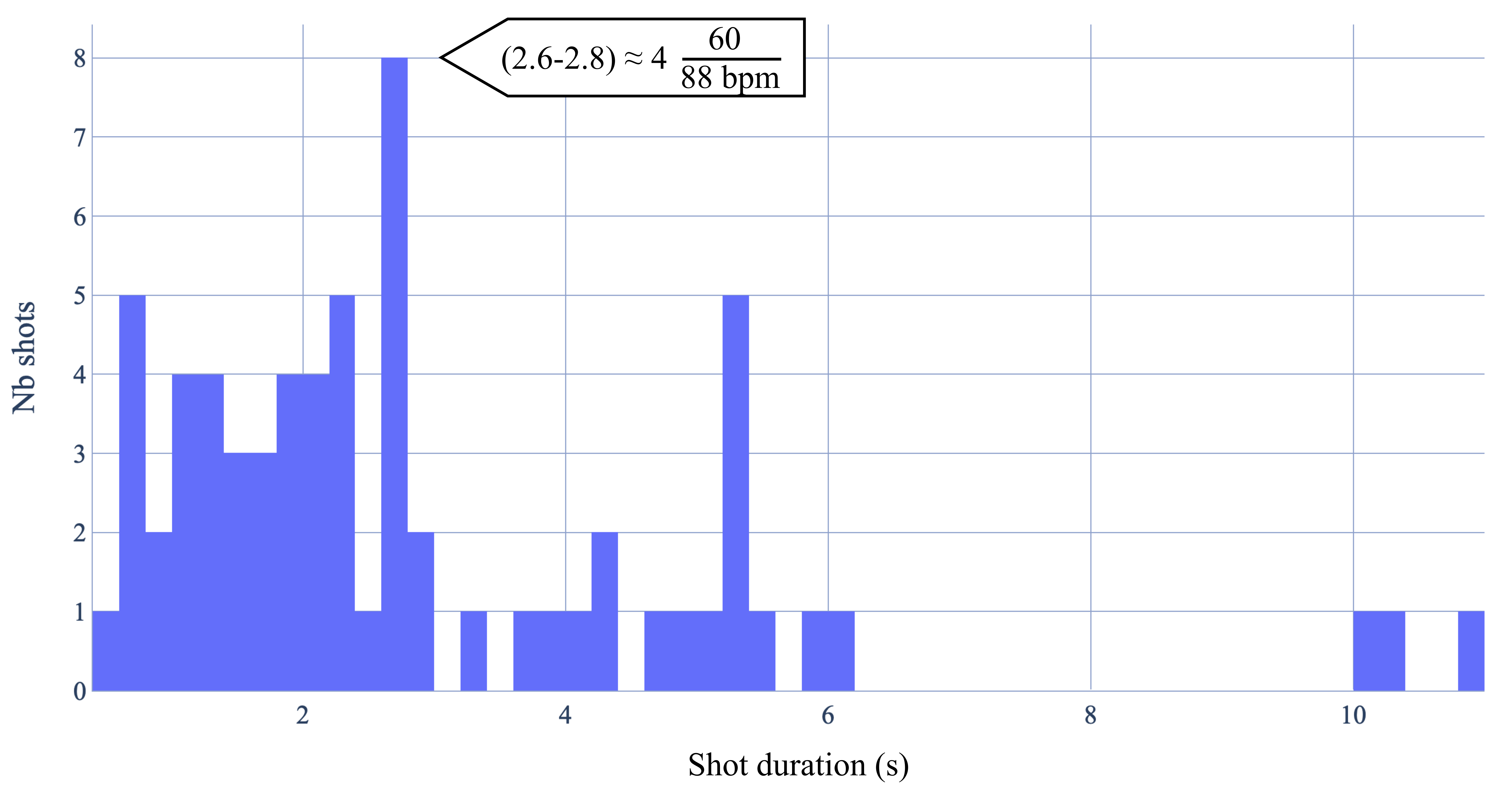}
    \end{subfigure}
    \begin{subfigure}{\columnwidth}
      \centering
        \includegraphics[width=0.9\columnwidth]{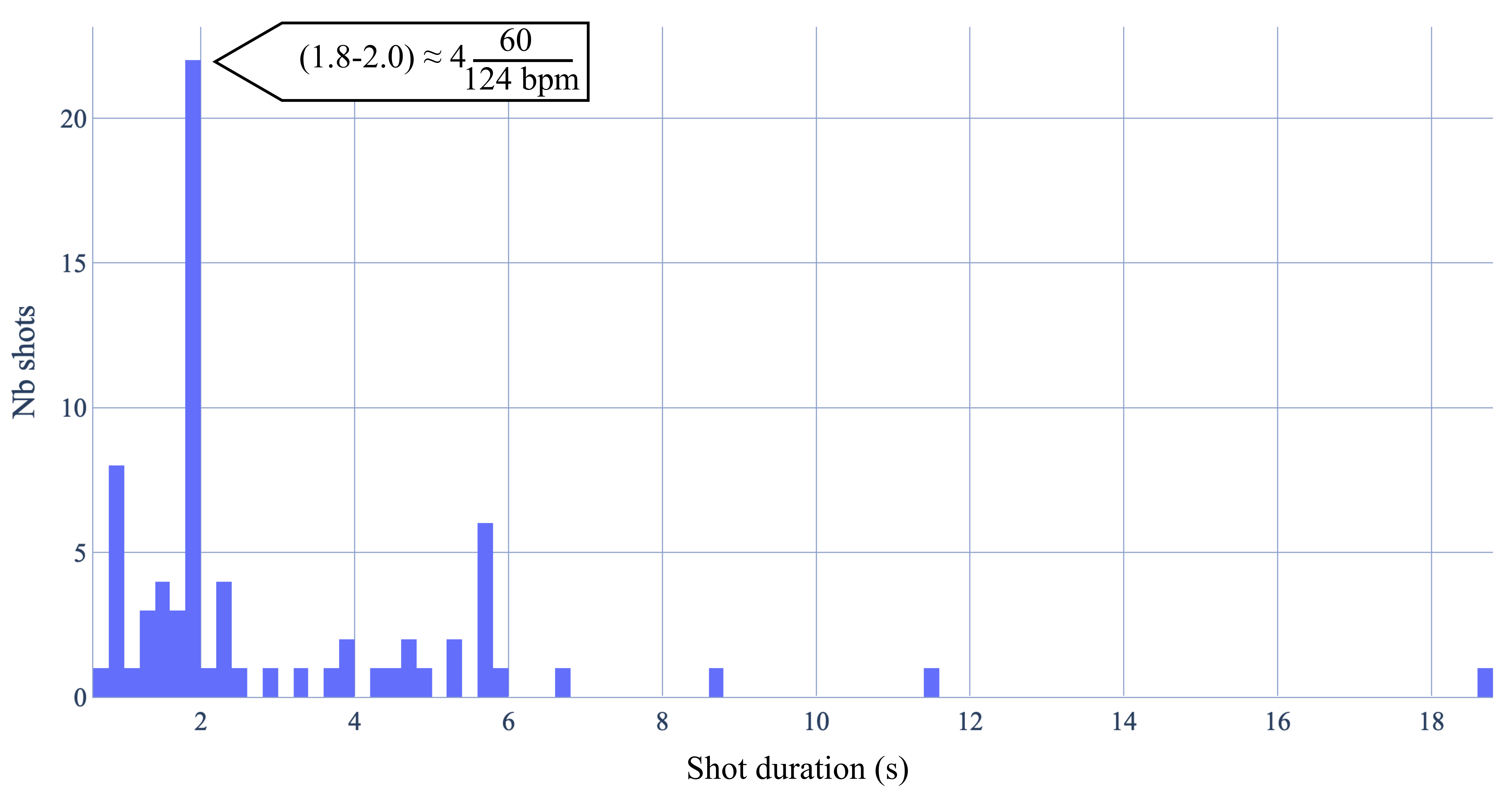}
    \end{subfigure}
    \caption{
    [top] Histogram of shot duration in the music video of \textit{Heartless} by Kanye West. The tempo is 88 bpm.
    [bottom] Histogram of shot duration in the music video of \textit{Firework} by Katy Perry. The tempo is 124 bpm.
    }
    \label{fig:histogram_cuts}
\end{figure}

This is for example the case of "Heartless" by Kanye West (see Figure \ref{fig:histogram_cuts} [top]) where the large peak at $\Delta_{shot}^{\max}$=2.72~s can be explained by the tempo at 88~bpm; or "Firework" by Katy Perry (see Figure \ref{fig:histogram_cuts} [bottom]) where the large peak at $\Delta_{shot}^{\max}$= 1.93~s can be explained by the tempo at 124~bpm. 

In our dataset, a synchronization at the bar level ($0.5\Delta_{bar} < \Delta_{shot}^{\max} < 1.5\Delta_{bar}$) occurs for one fifth of the clips (95 music videos).
Synchronization may also occur at other levels: at the beat level $\Delta_{beat}$, or the pattern level $\Delta_{pattern}$ (usually an even multiple of the bar duration).
In our dataset, a synchronization at the beat level ($0.5\Delta_{beat} < \Delta_{shot}^{\max} < 1.5\Delta_{beat}$) occurs for two thirds of the clips (329 music videos).
However, synchronization at pattern level $\Delta_{pattern} = 4 \Delta_{bar}$ almost never occurs (2 music videos).

\begin{table}
\footnotesize
    \caption{Agreement of musical structure (bar $\Delta_{bar}$ and beat $\Delta_{beat}$ level) and dominant shot duration $\Delta_{shot}^{\max}$ according to the \textbf{music genre} [top table] and according to the \textbf{video genre} [bottom table].
    Highest values are highlighted in bold, lowest values in italic.}
    \label{tab:shot_duration}
    \centering
    \begin{tabular}{|c|c|c|c|c|}
    \hline 
    \multirow{2}{*}{} & \multicolumn{2}{c|}{$\Delta_{shot}^{\max} \simeq \Delta_{bar}$} & \multicolumn{2}{c|}{$\Delta_{shot}^{\max} \simeq \Delta_{beat}$} \\
    \cline{2-5}
    Music Genre & \# tracks &  \% & \# tracks & \% \\ \hline 
Alternative & 2 & 8.3 & 19 & \textbf{79.2} \\
Country & 10 & \textbf{29.4} & 16 & 47.1 \\
Dance/Electro & 12 & \textbf{24.5} & 28 & 57.1 \\
Hip-Hop & 12 & 12.6 & 69 & 72.6 \\
Pop & 40 & 14.8 & 158 & \textit{58.3} \\
R\&B & 1 & \textit{5.3} & 13 & 68.4 \\
Reggaeton & 1 & 8.3 & 9 & \textbf{75.0} \\
Rock & 4 & \textbf{23.5} & 10 & \textit{58.8} \\
    \hline 
    \hline 
    \multirow{2}{*}{} & \multicolumn{2}{c|}{$\Delta_{shot}^{\max} \simeq \Delta_{bar}$} & \multicolumn{2}{c|}{$\Delta_{shot}^{\max} \simeq \Delta_{beat}$} \\
    \cline{2-5}
    Video Genre & \# tracks &  \% & \# tracks & \% \\ \hline
Concept & 33 & 14.5 & 148 & \textbf{65.2} \\
Dance & 11 & 17.7 & 40 & \textbf{64.5}\\
Narration & 28 & 17.5 & 93 & 58.1 \\
Performance & 19 & \textbf{25.7} & 41 & 55.4\\
Other & 4 & \textit{16.0} & 8 & \textit{32.0}\\
    \hline 
    \end{tabular}
\end{table}

In Table \ref{tab:shot_duration}, we indicate for each music genre and video genre, the number of tracks for which the $\Delta_{shot}^{\max}$ correspond to $\Delta_{bar}$ or $\Delta_{beat}$.
We only focus here on the most represented genres, i.e. which appear at least 10 times.
We observe a strong correspondence between $\Delta_{shot}^{\max}$ and $\Delta_{bar}$ for the music genres \texttt{Country}, \texttt{Dance/Electro} and \texttt{Rock} (one fourth of the tracks).
We observe a strong correspondence between $\Delta_{shot}^{\max}$ and $\Delta_{beat}$ for the music genres \texttt{Alternative} and \texttt{Reggaeton} (three quarters of the tracks).
This may imply, for example, that music video professionals favor more dynamic editing styles (using shorter shots on average) for \texttt{Reggaton} than for \texttt{Country} music.
We observe a strong correspondence between $\Delta_{shot}^{\max}$ and $\Delta_{bar}$ for the video genre \texttt{Performance} (one fourth of the tracks).
On the contrary, we observe a low correspondence between $\Delta_{shot}^{\max}$ and $\Delta_{beat}$ for the video genre \texttt{Other} (one third of the tracks).
It is likely that music videos in the \texttt{Other} category favor experimental  editing styles, with shots of more diverse duration.

As we see, there is a strong relationship between the video events and musical events duration.
This however does not mean that the changes occur simultaneously.
We study this in the next section.

\subsubsection{Comparison between events position}
\label{position}


Our second experiment aims at evaluating to what extent the musical events $l_{seg}$, $l_{bar}$ and video events $f_{shot}(t)$ happen simultaneously.
To measure this, we compute for each audio boundary $i$ ($t_i \in l_{seg}$ or $t_i \in l_{bar}$) a score $S_i \in [0,1]$. 
$S_i$ is defined as the integral over time of the shot boundary likelihood $f_{shot}(t)$ tampered by a non-normalized Gaussian window $w(t)$.
$w(t)$ is centered on $t_i$, with $\sigma=2$ (such that the effective duration of the window is approximately 0.5s at a frame rate of 25Hz) and with $w(0)=1$.
\begin{equation*}
    S_i = \int_t w(t-t_i) f_{shot}(t) dt, \;\;\; \forall t_i \in \{l_{seg}, l_{bar} \}
\end{equation*}
A large value of $S_i$ indicates that the $t_i$ position (the music structural event) corresponds to a large probability of shot boundary.
We then average $S_i$ for all audio boundaries $i$ to get $S$.
$S$ might be considered as a measure of precision, since it provides information on \textit{how many audio boundaries are explained by a video boundary}.
It should be noted that the number of video boundaries is larger than the number of audio boundaries (as seen in Figures~\ref{fig:firework} and  \ref{fig:rollinginthedeep}).
$S$ is also close to the measure proposed by \cite{Cemgil2000OnFiltering} to evaluate the performances of beat-tracking algorithms.
A large value of $S$ indicates that the shot boundaries are located at the same positions as the music structural events $l_{seg}$ or $l_{bar}$.
We compute $S$ separately using the $t_i$ from $l_{seg}$ or from $l_{bar}$.
To check if the amount of music-video event synchronization depends on the music and video genre, we average $S$ over all tracks of a given genre (music or video).

\begin{table}[t]
    \centering
    \footnotesize
        \caption{Shot transition intensity $S$ around music boundaries (either functional segments boundaries $l_{seg}$ or bar boundaries $l_{bar}$) according to \textbf{music genre} [top table] and according to the \textbf{video genre} [bottom table].
        Mean values and confidence intervals at 95\% are displayed.
    Highest values are highlighted in bold, lowest values in italic.}
    \label{tab:transnet_intensity}
    \begin{tabular}{|c|c|c|c|c|c|}
    \hline 
    Music Genre & $S(l_{seg})$ & $S(l_{bar})$ & \# tracks \\ \hline 
Alternative & 0.22 $\pm$ 0.08 & 0.23 $\pm$ 0.02 & 24 \\
Country & 0.20 $\pm$ 0.06 & 0.21 $\pm$ 0.02 & 34 \\
Dance/Electro & \textit{0.18 $\pm$ 0.05} & 0.21 $\pm$ 0.02 &  49 \\
Hip-Hop & 0.19 $\pm$ 0.03 & 0.25 $\pm$ 0.01 & 95  \\
Pop & \textbf{0.36 $\pm$ 0.02} & 0.21 $\pm$ 0.01 & 271  \\
R\&B & \textbf{0.29 $\pm$ 0.10} & \textbf{0.31 $\pm$ 0.03} & 19  \\
Reggaeton & 0.24 $\pm$ 0.11 & \textbf{0.28 $\pm$ 0.04} & 12  \\
Rock & \textit{0.18 $\pm$ 0.07} & \textit{0.19 $\pm$ 0.03} & 17  \\
    \hline 
    \hline 
    Video Genre & $S(l_{seg})$ & $S(l_{bar})$ & \# tracks \\ \hline 
Concept & 0.20 $\pm$ 0.02 & \textbf{0.23 $\pm$ 0.01} & 227 \\
Dance & \textbf{0.18 $\pm$ 0.04} & \textbf{0.24 $\pm$ 0.01} & 62 \\
Narration & \textbf{0.18 $\pm$ 0.03} & \textbf{0.23 $\pm$ 0.01} & 160 \\
Performance & \textit{0.15 $\pm$ 0.04} & 0.16 $\pm$ 0.01 & 74 \\
Other & \textit{0.11 $\pm$ 0.06} & \textit{0.11 $\pm$ 0.02} & 25 \\
    \hline 
    \end{tabular}
\end{table}

\textbf{Co-occurrence of music/video events by music genre.}
Table \ref{tab:transnet_intensity} [top part] shows the co-occurrence scores $S$ aggregated over music genres.
We observe variations of the values of $S$ according to the music genre.
For \texttt{Pop}, $S(l_{seg})$ is large (0.36) indicating that many shot transitions occur at the functional segment boundaries positions.
For \texttt{R\&B} and \texttt{Reggaeton}, $S(l_{bar})$ is large (0.31 and 0.28) indicating that many shot transitions occur at the downbeat positions.
We also observe that the value of $S(l_{seg})$ and $S(l_{bar})$ vary according to the music genre with very small values for \texttt{Dance/Electronic}, \texttt{Hip-Hop} and \texttt{Rock}.
This comes as a surprise especially for \texttt{Dance/Electronic}, because in the previous experiment, we observed a strong correspondence between the duration of shots and bars for this music genre. This shows that even though bars and shots have similar duration, their boundaries might not always co-occur.

\textbf{Co-occurrence of music/video events by video genre.}
Table \ref{tab:transnet_intensity} [bottom part] shows the co-occurrence scores $S$ aggregated over video genres.
We observe variations of the values of $S$ according to the video genre.
We see that the \texttt{Dance} video genre has a large value of $S(l_{bar})$ (0.24), which is not surprising given that video labeled as \texttt{Dance} actually show people dancing on the beat.
We also observe large values of $S(l_{bar})$ for the \texttt{Concept} and \texttt{Narration} video genres with consistent synchronization on the downbeats.
For the \texttt{Performance} video genre (the band is playing in front of the camera), we don't observe such a large correspondence ($S(l_{bar})=0.16$).
For the \texttt{Other} video genre, the low values ($S(l_{bar}) = S(l_{seg})=0.11$) are not surprising, given that some videos are very experimental and may feature complex video transitions, which may be difficult to detect by the TransNet.

\section{Conclusion}
\label{sec:conclusion}

According to the professionals and to our experiments, official music videos are edited by taking into account the music structure.
Although some experts mentioned that synchronization was often a matter of taste and intuition, we were able to bring out some trends.
We showed that the co-occurrence of music and video structural events would vary according to the music and video genres.
These elements can be reused to design or improve automatic music-video recommendation systems.
For example, if the task is to recommend an illustration video for a Pop or R\&B track, the system is expected to favor candidates that allow high synchronization of the structural events.

However, we have the intuition that other factors may impact the editing style of \acp{omv}.
In future work, we plan to investigate the role of other metadata, such as release date, artist popularity or harmonic complexity.
Although we focused on \acp{omv} for this study, we believe that a similar analysis can be conducted on other types of musical videos, e.g. movies or commercials.


\bibliography{references}

\begin{thebibliography}{10}
\providecommand{\url}[1]{#1}
\csname url@samestyle\endcsname
\providecommand{\newblock}{\relax}
\providecommand{\bibinfo}[2]{#2}
\providecommand{\BIBentrySTDinterwordspacing}{\spaceskip=0pt\relax}
\providecommand{\BIBentryALTinterwordstretchfactor}{4}
\providecommand{\BIBentryALTinterwordspacing}{\spaceskip=\fontdimen2\font plus
\BIBentryALTinterwordstretchfactor\fontdimen3\font minus
  \fontdimen4\font\relax}
\providecommand{\BIBforeignlanguage}[2]{{%
\expandafter\ifx\csname l@#1\endcsname\relax
\typeout{** WARNING: IEEEtran.bst: No hyphenation pattern has been}%
\typeout{** loaded for the language `#1'. Using the pattern for}%
\typeout{** the default language instead.}%
\else
\language=\csname l@#1\endcsname
\fi
#2}}
\providecommand{\BIBdecl}{\relax}
\BIBdecl

\bibitem{Arandjelovic2017LookLearn}
R.~Arandjelovic and A.~Zisserman, ``{Look, Listen and Learn},'' in
  \emph{Proceedings of IEEE ICASSP (International Conference on Computer
  Vision)}, Venice, Italy, 2017.

\bibitem{Gillet2007OnVideos}
O.~Gillet, S.~Essid, and G.~Richard, ``{On the correlation of automatic audio
  and visual segmentations of music videos},'' \emph{IEEE Transactions on
  Circuits and Systems for Video Technology}, vol.~17, no.~3, pp. 347--355,
  2007.

\bibitem{Wang2012TheVideo}
J.-C. Wang, Y.-H. Yang, I.-H. Jhuo, Y.-Y. Lin, and H.-M. Wang, ``{The
  acousticvisual emotion guassians model for automatic generation of music
  video},'' in \emph{Proceedings of ACM Multimedia}, Nara, Japan, 2012.

\bibitem{Kuo2013BackgroundAnalysis}
F.-F. Kuo, M.-K. Shan, and S.-Y. Lee, ``{Background Music Recommendation for
  Video Based on Multimodal Latent Semantic Analysis},'' in \emph{Proceedings
  of ICME (International Conference on Multimedia and Expo)}, San Jose, CA,
  USA, 2013.

\bibitem{Lin2015EMV-matchmaker:Generation}
J.~C. Lin, W.~L. Wei, and H.~M. Wang, ``{EMV-matchmaker: Emotional temporal
  course modeling and matching for automatic music video generation},'' in
  \emph{Proceedings of ACM Multimedia}, Brisbane, Australia, 2015, pp.
  899--902.

\bibitem{Lin2017AutomaticEditing}
J.~C. Lin, W.~L. Wei, J.~Yang, H.~M. Wang, and H.~Y.~M. Liao, ``{Automatic
  music video generation based on simultaneous soundtrack recommendation and
  video editing},'' in \emph{Proceedings of MMM (International Conference on
  Multimedia Modeling)}, Reykjavik, Iceland, 2017.

\bibitem{HUA2004AutomaticAnalysis}
X.-S. Hua, L.~Lu, and H.-J. Zhang, ``{Automatic music video generation based on
  temporal pattern analysis},'' in \emph{Proc. of ACM Multimedia}, New York
  City, NY, USA, 2004.

\bibitem{Wang2006FullyComposition}
J.~Wang, E.~Chng, and C.~Xu, ``{Fully and Semi-Automatic Music Sports Video
  Composition},'' in \emph{Proceedings of ICME (International Conference on
  Multimedia and Expo)}, Toronto, ON, Canada, 2006.

\bibitem{Wang2007GenerationCues}
J.~Wang, E.~Chng, C.~Xu, {Hanqinq Lu}, and Q.~Tian, ``{Generation of
  Personalized Music Sports Video Using Multimodal Cues},'' \emph{IEEE
  Transactions on Multimedia}, vol.~9, no.~3, 2007.

\bibitem{DownieMusicEXchange.}
\BIBentryALTinterwordspacing
J.~S. Downie, ``{Music Information Retrieval Evaluation eXchange.}'' [Online].
  Available: \url{http://www.music-ir.org/mirex/wiki/MIREX_HOME}
\BIBentrySTDinterwordspacing

\bibitem{Aner2002VideoClustering}
A.~Aner and J.~R. Kender, ``{Video Summaries through Mosaic-Based Shot and
  Scene Clustering},'' in \emph{Proceedings of ECCV (European Conference on
  Computer Vision)}, Copenhagen, Denmark, 2002, pp.~--.

\bibitem{Schindler2016HarnessingRetrieval}
A.~Schindler and A.~Rauber, ``{Harnessing music-related visual stereotypes for
  music information retrieval},'' \emph{ACM Transactions on Intelligent Systems
  and Technology}, vol.~8, no.~2, pp. 1--21, 2016.

\bibitem{Nieto2019TheMusic}
\BIBentryALTinterwordspacing
O.~Nieto, M.~Mccallum, M.~E.~P. Davies, A.~Robertson, A.~Stark, and E.~Egozy,
  ``{The Harmonix Set: Beats, Downbeats, and Functional Segment Annotations of
  Western Popular Music},'' in \emph{Proceedings of ISMIR (International
  Conference on Music Information Retrieval)}, Delft, The Netherlands, 2019.
  [Online]. Available:
  \url{https://archives.ismir.net/ismir2019/paper/000068.pdf}
\BIBentrySTDinterwordspacing

\bibitem{Inskip2008MusicRetrieval}
\BIBentryALTinterwordspacing
C.~Inskip, A.~Macfarlane, and P.~Rafferty, ``{Music, Movies and Meaning:
  Communication in Film-makers’ Search for Pre-existing Music, and the
  Implications for Music Information Retrieval},'' in \emph{Proceedings of
  ISMIR (International Conference on Music Information Retrieval)},
  Philadelphia, PA, USA, 2008. [Online]. Available:
  \url{https://archives.ismir.net/ismir2008/paper/000117.pdf}
\BIBentrySTDinterwordspacing

\bibitem{Gillet2007TranscriptionAudiovisuelles}
\BIBentryALTinterwordspacing
O.~Gillet, ``{Transcription des signaux percussifs. Application {\`{a}}
  l'analyse de sc{\`{e}}nes musicales audiovisuelles},'' Ph.D. dissertation,
  Ecole Nationale Superieure des Telecommunications, 2007. [Online]. Available:
  \url{https://pastel.archives-ouvertes.fr/pastel-00002805}
\BIBentrySTDinterwordspacing

\bibitem{RuiduoYang2004MusicObservation}
{Ruiduo Yang} and M.~Brown, ``{Music database query with video by synesthesia
  observation},'' in \emph{Proceedings of ICME (International Conference on
  Multimedia and Expo)}.\hskip 1em plus 0.5em minus 0.4em\relax Taipei, Taiwan:
  IEEE, 2004, pp.~--.

\bibitem{Mulhem2003PivotMixing}
P.~Mulhem, M.~S. Kankanhalli, J.~Yi, and H.~Hassan, ``{Pivot vector space
  approach for audio-video mixing},'' \emph{IEEE Multimedia}, vol.~10, no.~2,
  pp. 28--40, 4 2003.

\bibitem{Schindler2019Multi-ModalAnalysis}
\BIBentryALTinterwordspacing
A.~Schindler, ``{Multi-Modal Music Information Retrieval: Augmenting
  Audio-Analysis with Visual Computing for Improved Music Video Analysis},''
  Ph.D. dissertation, Technische Universit{\"{a}}t Wien, 2019. [Online].
  Available: \url{https://arxiv.org/abs/2002.00251}
\BIBentrySTDinterwordspacing

\bibitem{Schindler2019OnVideos}
\BIBentryALTinterwordspacing
A.~Schindler and A.~Rauber, ``{On the Unsolved Problem of Shot Boundary
  Detection for Music Videos},'' in \emph{Proceedings of MMM (International
  Conference on Multimedia Modeling)}, vol. 11295 LNCS.\hskip 1em plus 0.5em
  minus 0.4em\relax Thessaloniki, Greece: Springer Verlag, 2019. [Online].
  Available:
  \url{https://link.springer.com/chapter/10.1007%2F978-3-030-05710-7_43}
\BIBentrySTDinterwordspacing

\bibitem{Jia2019DeepReview}
B.~Jia, J.~Lv, and D.~Liu, ``{Deep learning-based automatic downbeat tracking:
  a brief review},'' \emph{Multimedia Systems}, vol.~25, no.~6, 12 2019.

\bibitem{Bock2016Madmom:Library}
S.~B{\"{o}}ck, F.~Korzeniowski, J.~Schl{\"{u}}ter, F.~Krebs, and G.~Widmer,
  ``{Madmom: A new Python audio and music signal processing library},'' in
  \emph{Proceedings of ACM Multimedia}, New York City, NY, USA, 2016, pp.
  1174--1178.

\bibitem{Bogdanov2013ESSENTIA:Analysis}
D.~Bogdanov, X.~Serra, N.~Wack, E.~G{\'{o}}mez, S.~Gulati, P.~Herrera,
  O.~Mayor, G.~Roma, J.~Salamon, and J.~Zapata, ``{ESSENTIA: an Open-Source
  Libraryfor Sound and Music Analysis},'' in \emph{Proceedings of ACM
  Multimedia}.\hskip 1em plus 0.5em minus 0.4em\relax New York, NY, USA: ACM
  Press, 2013.

\bibitem{Bock2016JointNetworks.}
\BIBentryALTinterwordspacing
S.~B{\"{o}}ck, F.~Krebs, and G.~Widmer, ``{Joint Beat and Downbeat Tracking
  with Recurrent Neural Networks.}'' in \emph{Proceedings of ISMIR
  (International Society for Music Information Retrieval Conference)}, New York
  City, NY, USA, 2016. [Online]. Available:
  \url{https://archives.ismir.net/ismir2016/paper/000186.pdf}
\BIBentrySTDinterwordspacing

\bibitem{McFee2014LearningAnalysis}
B.~McFee and D.~P. Ellis, ``{Learning to segment songs with ordinal linear
  discriminant analysis},'' in \emph{Proceedings of IEEE ICASSP (International
  Conference on Acoustics, Speech and Signal Processing)}, Florence, Italy,
  2014, pp. 5197--5201.

\bibitem{Soucek2019TransNet:Transitions}
\BIBentryALTinterwordspacing
T.~Sou{\v{c}}ek, J.~Moravec, and J.~Loko{\v{c}}, ``{TransNet: A deep network
  for fast detection of common shot transitions},'' \emph{arXiv preprint
  arXiv:1906.03363}, 2019. [Online]. Available:
  \url{https://arxiv.org/abs/1906.03363}
\BIBentrySTDinterwordspacing

\bibitem{Nieto2016SystematicResearch.}
\BIBentryALTinterwordspacing
O.~Nieto and J.~P. Bello, ``{Systematic Exploration Of Computational Music
  Structure Research.}'' in \emph{Proceedings of ISMIR (International
  Conference on Music Information Retrieval)}, New York City, NY, USA, 2016.
  [Online]. Available:
  \url{https://archives.ismir.net/ismir2016/paper/000043.pdf}
\BIBentrySTDinterwordspacing

\bibitem{Hennequin2018AudioTags}
\BIBentryALTinterwordspacing
R.~Hennequin, J.~Royo-Letelier, and M.~R. Moussallam~Deezer, ``{Audio Based
  Disambiguation of Music Genre Tags},'' in \emph{Proceedings of ISMIR
  (International Conference on Music Information Retrieval)}, 2018. [Online].
  Available: \url{https://archives.ismir.net/ismir2018/paper/000163.pdf}
\BIBentrySTDinterwordspacing

\bibitem{Defferrard2018LearningAudio}
M.~Defferrard, S.~P. Mohanty, S.~F. Carroll, and M.~Salath{\'{e}}, ``{Learning
  to Recognize Musical Genre from Audio},'' in \emph{Companion of WWW (The Web
  Conference)}, New York City, NY, USA, 2018, pp.~--.

\bibitem{Bertin-Mahieux2011TheDataset}
\BIBentryALTinterwordspacing
T.~Bertin-Mahieux, D.~P. Ellis, B.~Whitman, and P.~Lamere, ``{The Million Song
  Dataset},'' in \emph{Proceedings of ISMIR (International Conference on Music
  Information Retrieval)}, Miami, FL, USA, 2011. [Online]. Available:
  \url{https://ismir2011.ismir.net/papers/OS6-1.pdf}
\BIBentrySTDinterwordspacing

\bibitem{Gemmeke2017AudioEvents}
J.~F. Gemmeke, D.~P.~W. Ellis, D.~Freedman, A.~Jansen, W.~Lawrence,
  R.~Channing~Moore, M.~Plakal, and M.~Ritter, ``{Audio Set: An Ontology and
  Human-Labeled Dataset for Audio Events},'' in \emph{Proceedings of IEEE
  ICASSP (International Conference on Acoustics, Speech and Signal
  Processing)}, New Orleans, LA, USA, 2017.

\bibitem{Defferrard2017FMA:Analysis}
\BIBentryALTinterwordspacing
M.~Defferrard, K.~Benzi, P.~Vandergheynst, and X.~Bresson, ``{FMA: A Dataset
  for Music Analysis},'' in \emph{Proceedings of ISMIR (International Society
  for Music Information Retrieval Conference)}, Suzhou, China, 2017. [Online].
  Available: \url{https://archives.ismir.net/ismir2017/paper/000075.pdf}
\BIBentrySTDinterwordspacing

\bibitem{You2010AAnalysis}
J.~You, G.~Liu, and A.~Perkis, ``{A semantic framework for video genre
  classification and event analysis},'' \emph{Signal Processing: Image
  Communication}, vol.~25, no.~4, 2010.

\bibitem{Varghese2019ARelevance}
\BIBentryALTinterwordspacing
J.~Varghese and K.~N. Ramachandran~Nair, ``{A Novel Video Genre Classification
  Algorithm by Keyframe Relevance},'' in \emph{Information and Communication
  Technology for Intelligent Systems. Smart Innovation, Systems and
  Technologies}.\hskip 1em plus 0.5em minus 0.4em\relax Singapore: Springer,
  2019, vol. 106. [Online]. Available:
  \url{https://link.springer.com/chapter/10.1007%2F978-981-13-1742-2_68}
\BIBentrySTDinterwordspacing

\bibitem{Choros2018VideoShots}
\BIBentryALTinterwordspacing
K.~Choro{\'{s}}, ``{Video Genre Classification Based on Length Analysis of
  Temporally Aggregated Video Shots},'' in \emph{Computational Collective
  Intelligence. Lecture Notes in Computer Science}.\hskip 1em plus 0.5em minus
  0.4em\relax Springer, Cham, 2018, vol. 11056. [Online]. Available:
  \url{https://link.springer.com/chapter/10.1007%2F978-3-319-98446-9_48}
\BIBentrySTDinterwordspacing

\bibitem{Cemgil2000OnFiltering}
\BIBentryALTinterwordspacing
A.~T. Cemgil, B.~Kappen, P.~Desain, and H.~Honing, ``{On tempo tracking:
  Tempogram representation and Kalman filtering},'' \emph{Journal of New Music
  Research}, vol.~29, no.~4, pp. 259--273, 2000. [Online]. Available:
  \url{https://www.mcg.uva.nl/papers/mmm-26.pdf}
\BIBentrySTDinterwordspacing

\end{thebibliography}


\end{document}